\documentclass[twocolumn,showpacs,aps,prb]{revtex4}
\usepackage{stmaryrd}
\usepackage{amssymb}
\usepackage{graphicx}
\usepackage{dcolumn}
\usepackage{bm}
\usepackage{color}

\begin{document}
\preprint{Physical Review B} %

\title{Aharonov-Bohm phase operations on a double-barrier nanoring charge qubit}
\author{Yuquan Wang}
\affiliation{%
Department of Physics, Tsinghua University, Beijing 100084, People's
Republic of China}
\author{Ning Yang}
\affiliation{%
Department of Physics, Tsinghua University, Beijing 100084, People's
Republic of China}
\author{Jia-Lin Zhu}
\email[Email address: ]{zjl-dmp@tsinghua.edu.cn}
\affiliation{%
Department of Physics, Tsinghua University, Beijing 100084,
People's Republic of China}

\date{\today}

\begin{abstract}
We present a scheme for charge qubit implementation in a
double-barrier nanoring. The logical states of the qubit are encoded
in the spatial wavefunctions of the two lowest energy states of the
system. The Aharonov-Bohm phase introduced by magnetic flux, instead
of tunable tunnelings, along with electric fields can be used for
implementing the quantum gate operations. During the operations, the
external fields should be switched smoothly enough to avoid the
errors caused by the transition to higher-lying states. The
structure and field effects on the validity of the qubit are also
studied.
\end{abstract}
\maketitle
\section{Introduction}
Solid state systems seem to be good candidates for quantum computing
implementations. In some schemes, qubits can be encoded in nuclear
(or electron) spin states.\cite{Kane1998, Golovach2002} Although
spin states have relatively long decoherence times ($\sim ms$), spin
operations are rather slow processes. Besides, the single-spin
measurements\cite{Elzerman2004, Greentree2005} are still a great
challenge. Many recent researches focus on charge-based quantum
computing technology,\cite{Hollenberg2004, Openov2004, Barrett2006}
in which qubits are encoded in the charge degrees of freedom.
Charge-based qubits have shorter decoherence times ($\sim ns$ for
GaAs\cite{Hayashi2003} and maybe much longer in some other
materials\cite{Gorman2005}) than their spin-based counterparts. Yet
all quantum gates can be done in very short times ($\sim ps$), which
are well below the decoherence times. The initialization and readout
of charge qubits have been proposed.

For charge-qubit implementation, a typical scheme is a coupled
quantum dots (CQDs) system with an electron tunneling back and
forth.\cite{Gorman2005} It can be easily scaled up based on the
staggered CPHASE or CNOT configuration.\cite{Hollenberg2004} Such
scheme, however, is based on the variability of the
tunneling.\cite{Fedichkin2000, Weichselbaum2004, Gorman2005} In many
physical quantum systems, the handle on the tunneling is limited or
impossible. Then the implementation of full single-qubit
manipulation requires tunable external magnetic fields and the
architecture of the system must be elaborately
designed.\cite{Weichselbaum2004, Ulloa2004}

Nowadays, benefiting from new fabrication and experiment techniques
we can fabricate ringlike quantum dot, namely nanoring, with various
materials, such as InAs/GaAs,\cite{Lorke2000, Fuhrer2001,
Keyser2002, Granados2003} Si,\cite{Hobbs2004, JHHe2005}
SiGe,\cite{Jiang2003, JHHe2004} and so on\cite{Kong2004}. Its
ringlike geometry is suitable for observing Aharonov-Bohm (AB)
effect. It has been shown that additional structures, such as two
barriers or impurities can bring unique electronic and transport
characters to the system. There are two important modes of AB
oscillations in double-barrier nanoring, named X and O
modes.\cite{Zhu2003} The ground state entanglement with
environment\cite{Cedraschi2001, Buttiker2005} and the persistent
current oscillations\cite{Buttiker2005, Buttiker1996} are also
widely studied. To a certain extent, such a system can be viewed as
CQDs with a multiply connected domain. Then it may serve as a charge
qubit and facilitate the quantum operation by changing AB phase
caused by the magnetic flux.

Employing the external fields to modify the wavefuntions of an
electron in nanostructures is the foundation of solid quantum
computation. So in this work, we will study the evolutions of the
wavefuntions and demonstrate the validity of the charge qubit based
on the double-barrier nanoring. Full single-qubit operations can be
carried out by electric fields and magnetic flux. The remainder of
this paper is organized as follows. The descriptions of model
Hamiltonian and the calculation method for the evolution of states
are presented in Sec.II. Main results and discussions are given in
Sec.III, followed by a summary in Sec.IV.

\date{\today}

\maketitle
\section{\protect\smallskip MODEL SYSTEM}
The model Hamiltonian for an electron in a two-dimensional ring with
two identically sectorial barriers, subjected to a magnetic flux
$\phi$ and an in-plane electric field F applied along the axis
$\theta=0$, as shown in Fig.1, is written as
\begin{equation}
H=\left ( -i\nabla + \frac {\phi} {|\bf r|} \right
)^2+V_{c}+V_{g}+{\bf F}\cdot {\bf r} \label{eq01}
\end{equation}
where $r_a$ and $r_b$ are , respectively, the inner and outer
radii. $V_c$ is the hard wall potential which is 0 in the ring and
infinite elsewhere. $V_g$ are the barriers in the ring whose
height are $V_0$ in the barriers and 0 elsewhere. The width of the
barrier is selected quite small to ensure that the wave function
has maximally one angular node inside each barrier.Here we use the
effective atomic units, in which the effective Rydberg
Ry$^{\text{*} }=m_e^{*}e^4/2\hbar ^2\left( 4\pi \varepsilon
_0\varepsilon _r\right) ^2$, the effective Bohr radius
$a_B^{*}=4\pi \varepsilon _0\varepsilon _r\hbar ^2/m_e^{*}e^2$ and
$\phi_0=2\pi\hbar c/e$ are taken to be the energy, length, and
magnetic flux units, respectively. The units for the electric
field is ${\text F}_0=$Ry$^{\text{*}}/ea_B^{*}$. For InAs/GaAs
materials, for example, Ry$^{\text{*}}=5.8$ meV, $a_B^{*}=10$ nm
and ${\text F}_0=5.8$ KV/cm. $\phi_0$ included by a 1D ring with a
radius of 10nm corresponds to the magnetic field $13.18$T.

\begin{figure}[ht]
\includegraphics*[angle=0,width=0.32\textwidth]{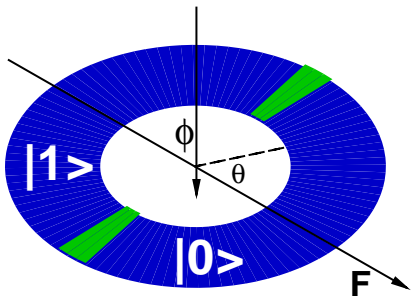}\\
\caption{\label{FIG:SchemeRing}(Color online) Scheme of a parallel
double-barrier nanoring subject to a magnetic flux and an in-plane
electric field.}
\end{figure}

The eigenenergies {$E_n$} and corresponding eigenstates {$\psi_n$}
of the double-barrier nanoring were computed by direct
diagonalization of the Hamiltonian in a basis set of 59 eigenstates
of the same nanoring without a barrier.\cite{Zhu2003} Since the
nanoring is divided into two segments by the barriers, the electron
states localized in right and left segments can be respectively
regarded as the logical state $|0\rangle$ and $|1\rangle$, as will
be shown below.

In order to demonstrate the implementation of quantum gate
operations, we need to calculate the evolutions of the states when
$\bf F$ and $\phi$ are changing. Assume that the Hamiltonian is the
sum of $H$ and $H'(t)$, where $H'(t)$ is the change of the
Hamiltonian from time $t=0$, the time evolution of a state
\begin{equation}
\psi(t)=\sum_k C_k(t)\exp \left (-\frac {i}{\hbar} E_k t \right
)\psi_k
\end{equation}
is given by
\[
i\hbar\frac {dC_n(t)}{dt} = \sum_k
C_k(t)\langle\psi_n|H'(t)|\psi_k\rangle\exp \left [ \frac {i}{\hbar}
(E_n-E_k)t \right ]
\]
This ordinary differential equation set were solved by Runge-Kutta
method. Then we can explore the external field and structure effects
in implementations of different gate operations.

\maketitle
\section{\protect\smallskip Analysis}
For the nanoring with a hard-wall potential, we can respectively
define the radius $R$ and the width $W$ as the average value of
inner and outer radii and the difference between them.

\subsection{Energy spectra and qubit definition}
\begin{figure}[ht]
\includegraphics*[angle=0, width=0.48\textwidth]{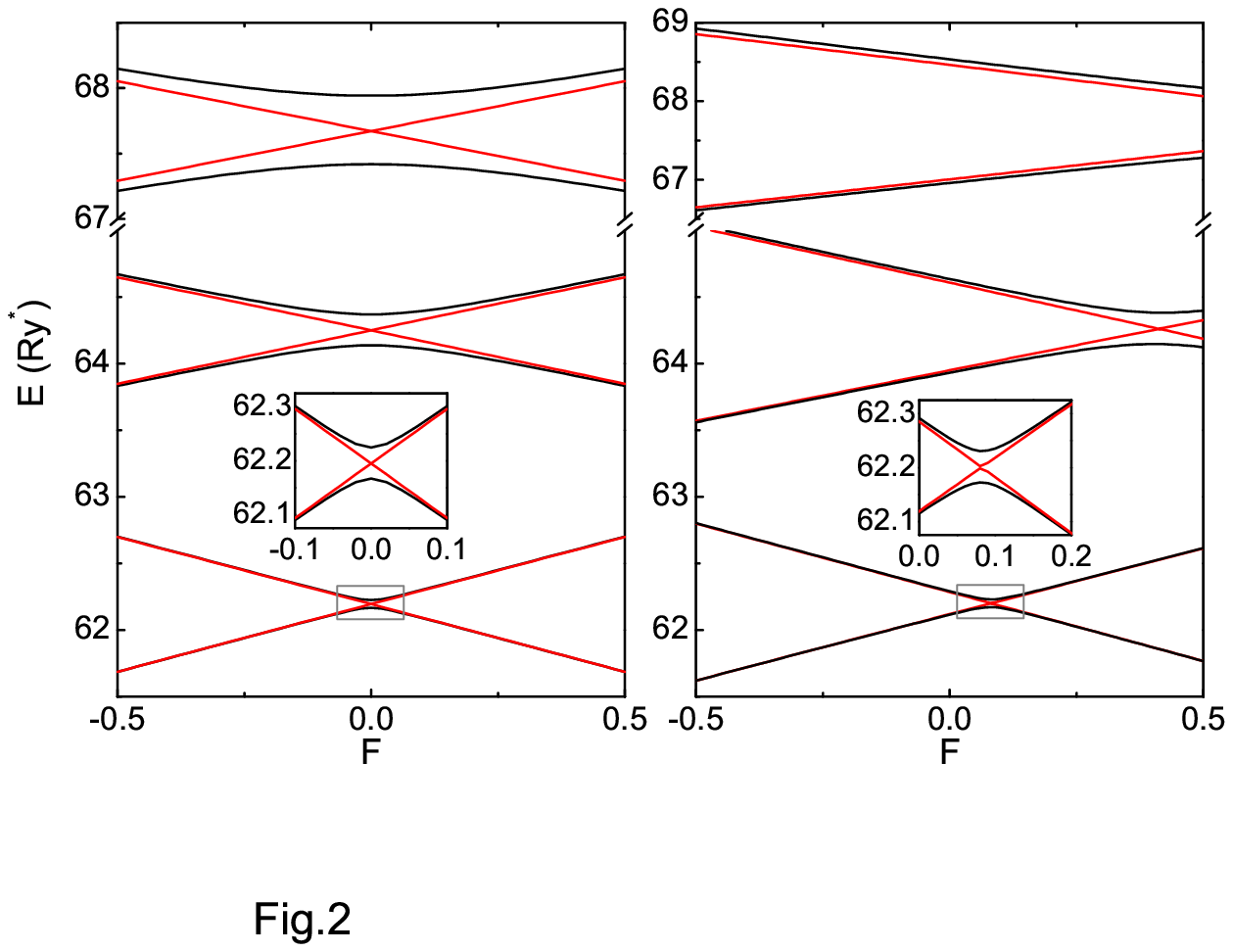}\\
\caption{\label{FIG:Ephi}(Color online) $\bf F$-dependence of energy
spectra of the nanoring with two parallel barriers (a) and
nonparallel barriers (b) with $R=1.2a_B^*$, $W=0.4a_B^*$,
$V_0=90$Ry$^{\text{*}}$, $\phi=0$(black lines) and $0.5\phi_0$(red
lines), respectively. The first two levels are also shown in the
inset for clarity.}
\end{figure}

In our scheme of charge qubit, the logical states are encoded in the
wavefunctions of the first two states of the double-barrier
nanoring. Due to the AB effect, it can be seen from the energy
spectra in Fig.2(a) that the ground state ($\psi_1$) and first
excited state ($\psi_2$) of a parallel double-barrier nanoring (the
angle between two barriers is equal to $\pi$) are degenerate when
$\phi=0.5\phi_0$ and ${\bf F}=0$. These two occasionally degenerate
states are just X-type states which had been discussed in previous
work.\cite{Zhu2003} The virtue of this condition is that the qubit
can be frozen in any linear superposition in its qubit space to
avoid unnecessary evolutions in idle time. This request can be also
satisfied by applying high enough barriers, just like the situation
in CQDs system. It is convenient to take the normalized sum and
difference of the two states $(\psi_1\pm \psi_2)/\sqrt{2}$ as
logical states $|0\rangle$ and $|1\rangle$, respectively. Within
such choice, the electron of state $|0\rangle$ ($|1\rangle$) is
almost completely localized in the right (left) segment of the ring.
The energy spectra for a nonparallel double-barrier nanoring are
shown in Fig.2(b). The first two energy levels can be still tuned to
degenerate by $\bf F$ with $\phi=0.5\phi_0$. So the qubit states can
be always defined properly to avoid needless evolutions. Due to the
inherent ringlike geometry of the system, in our charge qubit scheme
it is much easier to implement AB phase operation than that in CQDs.
In the following of the paper, we will focus on a parallel
double-barrier nanoring with a initial magnetic flux
$\phi=0.5\phi_0$. The initialization of qubit states to the state
$|0\rangle$ can be realized by applying an appropriate electric
field along the axis $\theta=0$. The readout can be implemented by a
single-electron transistor (SET)\cite{Aassime2001} or quantum point
contact (QPC)\cite{Buks1998, Pilgram2002, Gurvitz2003} detector.
Because the double-barrier nanoring can be also viewed as a new kind
of CQDs, the qubit based on it can be scaled up by present CPHASE
configuration scheme.\cite{Hollenberg2004}

\subsection{State evolutions with $\bf F$ and $\phi$}
\begin{figure}[ht]
\includegraphics*[angle=0, width=0.48\textwidth]{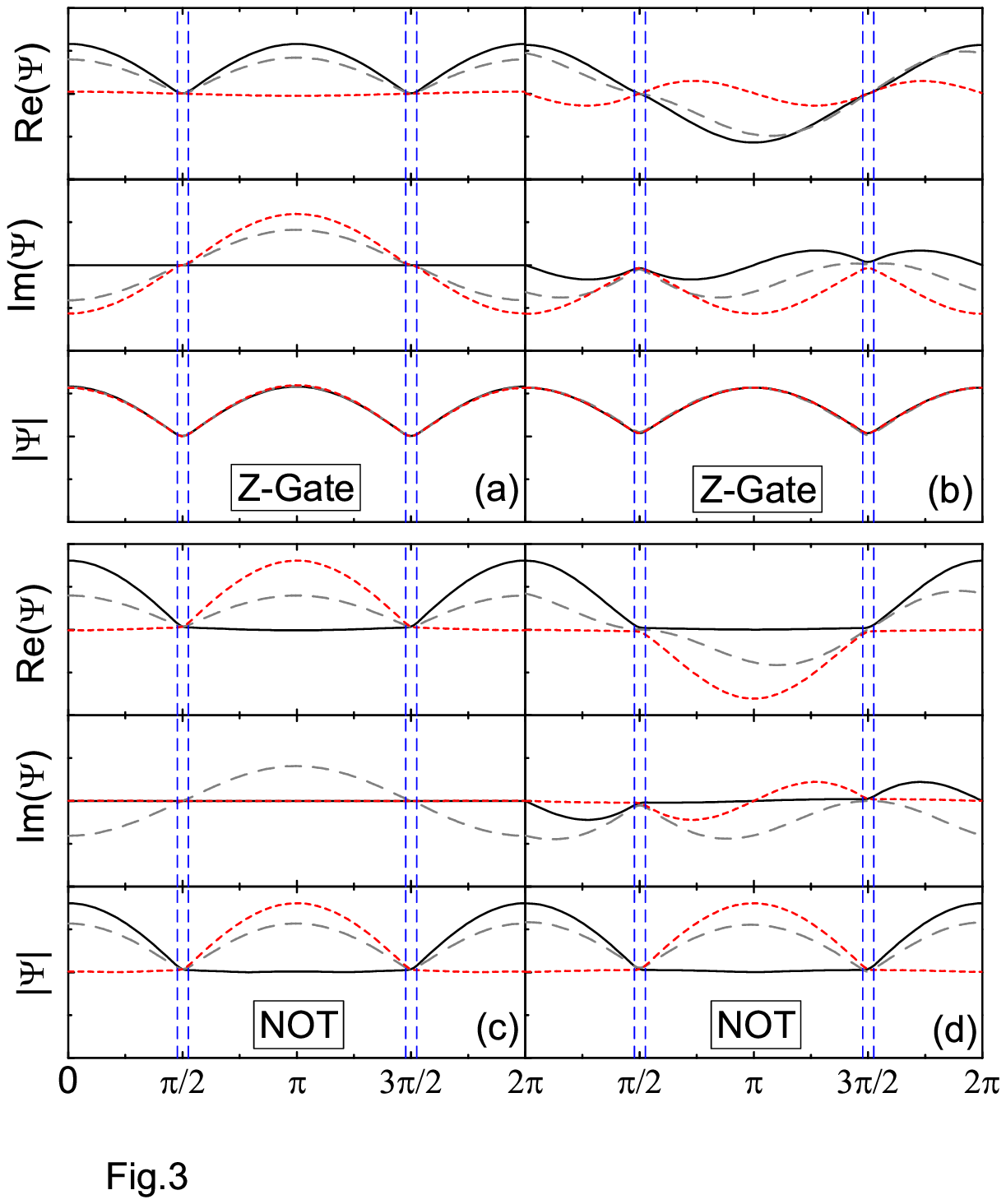}\\
\caption{\label{FIG:Ephi}(Color online) The evolutions of
wavefunctions during the Z-gate operations by applying $\bf F$ with
$V_0\rightarrow\infty$ and $\phi=0$ (a) or finite $V_0$ and
$\phi=0.5\phi_0$ (b), the NOT-gate operations by changing $V_0$ with
$\phi=0$ (c) or changing $\phi$ with fixed finite $V_0$ (d). The
real and imaginary parts and the modulus of the wavefunctions are
plotted in three rows respectively. Different line styles represent
the quantities at different time. The black solid lines and red
dotted lines represent the quantities at the starting and ending
time, respevtively. The vertical dashed lines indicate the positions
of the two barriers.}
\end{figure}

\begin{figure}[ht]
\includegraphics*[angle=0, width=0.45\textwidth]{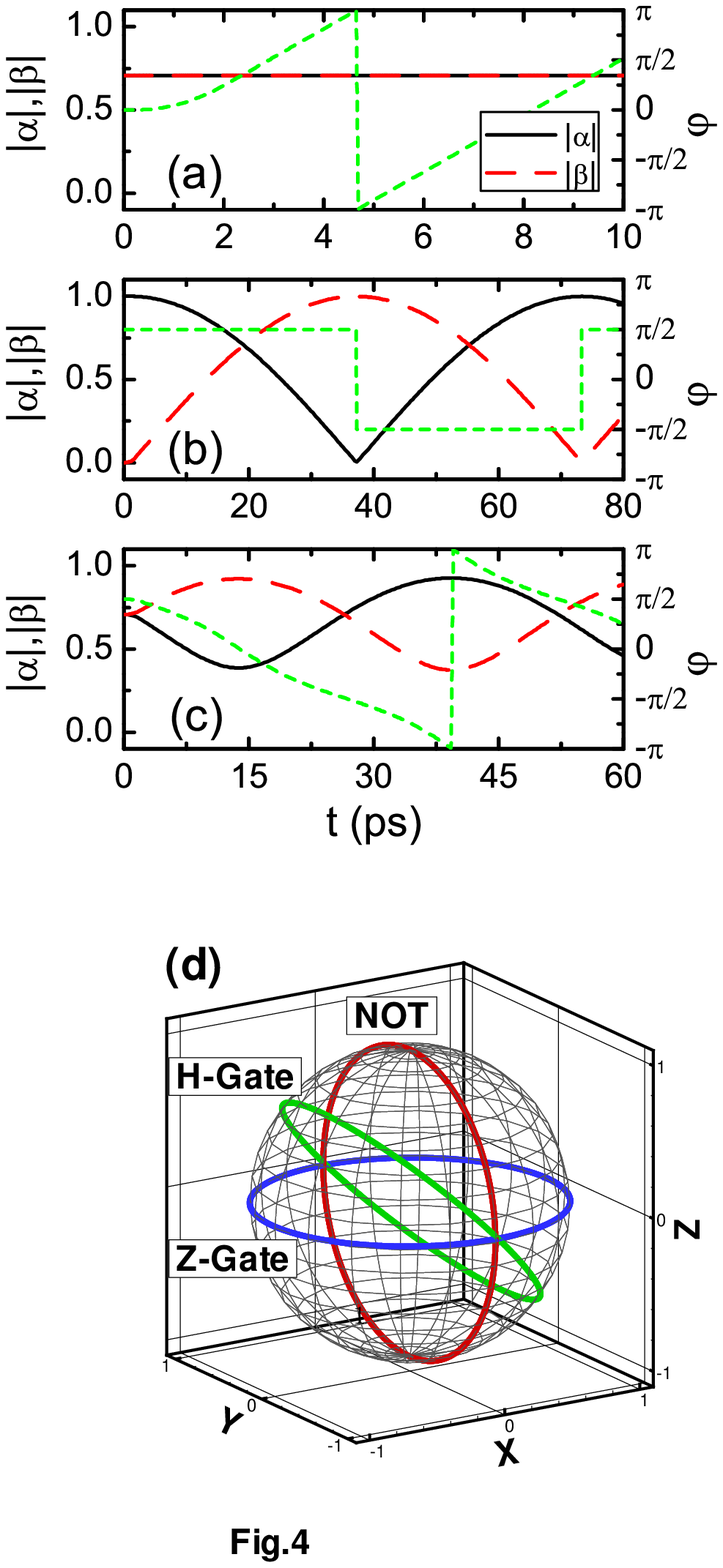}\\
\caption{\label{FIG:Ephi}(Color online) Evolutions of the qubit
states with in-plane electric field pulse (a), magnetic flux pulse
(b) and both fields together (c). Black solid, red dashed and green
dotted lines correspond to the square root of probability density of
the state occupying $|0\rangle$, $|1\rangle$ and the phase
difference, respectively. The trajectories of Bloch vectors
corresponding the evolutions in (a)$\sim$(c) are shown in (d) noted
as Z-, NOT- and H-gate, respectively.}
\end{figure}

\begin{figure*}[ht]
\includegraphics*[angle=0,width=0.98\textwidth]{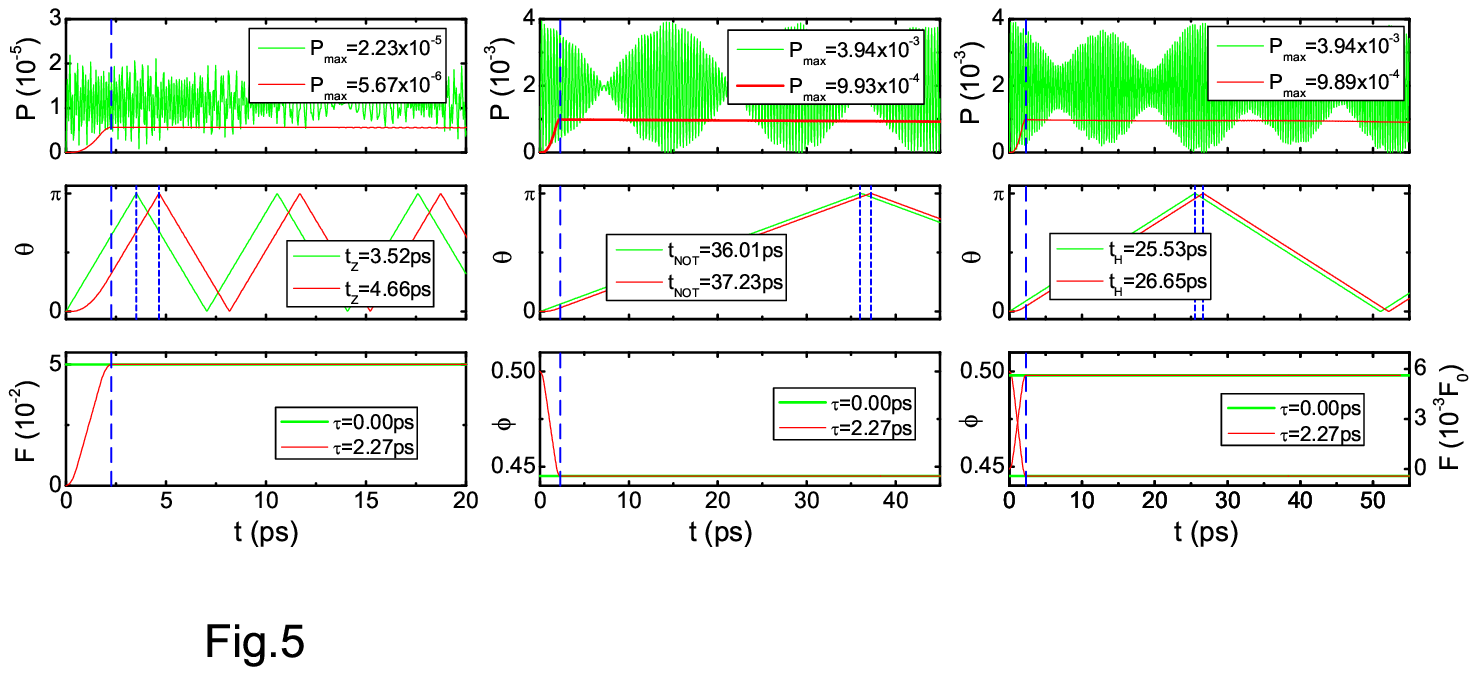}\\
\caption{\label{FIG:Operation}(Color online) Control parameter
settings (the bottom row), angular evolutions (the middle row) of
Bloch vectors in the plane of their trajectories and probabilities
(the upper row) lost from the qubit space to higher states under the
operation of the Z- (left column), NOT- (middle column) and Hadamard
gate (right column). The size of the ring is $R=1.2a_B^*$,
$W=0.4a_B^*$ and $V_0=90$Ry$^{\text{*}}$. The red and green lines
correspond to different pulsing form.}
\end{figure*}
With logical states $|0\rangle$ and $|1\rangle$ defined, we can
calculate the evolutions of the wavefunctions and qubit states with
the change of external parameters by employing Runge-Kutta method.
We will present the results concentrating on the characters of
evolutions in this subsection. The quantitative analyses are left to
next subsection.

From Hamiltonian (1) it can be imagined that changing the magnetic
flux, the height of the two barriers and the electric field all can
achieve some evolutions of wavefunctions. In Fig.3(a) and (b) we
illustrate the evolutions of wavefunctions when applying an in-plane
electric field along $\theta=0$ from $t=0$. The positions of the
barriers are indicated by the vertical dashed lines. The initial
state are both $\psi_1$, although the wavefunctions are not same
because $V_g$ is infinite and $\phi=0$ in Fig.3(a) but
$\phi=0.5\phi_0$ and $V_g$ is finite in Fig.3(b). Both conditions
make the states $\psi_1$ and $\psi_2$ degenerate. After some time
the final states can be both $\psi_2$, although their wavefunctions
are not equal either. Recalling the definition of logical states of
the qubit, the initial and final states are just
$(|0\rangle+|1\rangle)/\sqrt{2}$ and
$(|0\rangle-|1\rangle)/\sqrt{2}$, respectively. It can be seen from
the third row of Fig.3(a) and (b) that the modulus of the
wavefunctions are unchanged during the evolutions. This evolution is
just the single-qubit gate operation namely the Z-gate, which will
be discussed in the following section.

One goal of the paper is to identify the similar effects of changing
AB phase and the barrier heights on the qubit operations. Thus we
verify the evolutions of wavefuntions when changing the barrier
heights and the magnetic flux from $t=0$. The results are shown in
Fig.3(c) and (d). The initial states are both logical state
$|0\rangle$ with different wavefunctions, because there is no flux
in Fig.3(a) but $0.5\phi_0$ flux before $t=0$ in Fig.3(b). It can be
found that the final states are both $|1\rangle$. It means that the
two methods can implement similar quantum operations on the logical
states of the qubit. From Fig.2 it can be seen that the crossing of
the first two states becomes anticrossing when $\phi\neq0.5\phi_0$
which means that the two states can mix up. Similar with varying the
barrier height, the change of the flux determines the admixture
level. So in the following, we will adopt the magnetic flux, instead
of the barrier height, and the electric field to explore the
evolutions of the qubit states.

The first case which we concern is still the electric field. An
in-plane electric field applied along the axis $\theta=0$ from $t=0$
make the energies of one electron localized in the left and right
segments unequal. As shown in Fig.3(b) and Fig.4(a), the electric
field will not change the probability of the states projecting to
$|0\rangle$ or $|1\rangle$. It can only bring a phase difference
between the two components. The trajectory of Bloch vector
correspond to such an evolution is a rotation around z-axis of Bloch
sphere.

The situation is different when the magnetic flux changes. Because
we have assumed that there is an always-on magnetic flux
$\phi=0.5\phi_0$, we will decrease this flux from $t=0$ and the
initial state is chosen to be $|0\rangle$ without electric fields.
Then the degeneracy of $|0\rangle$ and $|1\rangle$ is removed, and
the mixture of the two states forms a cycle trajectory of Bloch
vector around x-axis of Bloch sphere. From Fig.4(b) we can see that
the variation of the flux changes the probability of each component.
It also changes the sign of the phase difference when Bloch vector
rotates half a cycle.

By applying both the electric field and magnetic flux pulse, we can
implement arbitrary rotation of Bloch vector. In Fig.4(c), we plot
one of such rotations which can be used for implementing the
Hadamard gate.

\subsection{Characters of quantum operations}
Through the above analysis, it is straightforward to implement the
single-qubit quantum operations by applying the electric field and
changing the magnetic flux.

The first important gate operation is the Z-gate which implements
the transform from $\alpha|0\rangle+\beta|1\rangle$ to
$\alpha|0\rangle-\beta|1\rangle$, where $\alpha$ and $\beta$ are
arbitrary constants. Applying an electric field and controlling the
pulse duration to make the angle of rotation equal to $\pi$ just
achieves a Z-gate operation. As shown in the left column of Fig.5,
for the nanoring subjected to a magnetic flux $\phi=0.5\phi_0$ with
$R=1.2a_B^*$, $W=0.4a_B^*$ and $V_0=90$Ry$^{\text{*}}$, the time for
completing a Z-gate operation is $4.66ps$ if the amplitude of the
electric pulse is $0.05{\text F}_0$. This operation time is much
smaller than the typical charge decoherence time for GaAs.

In order to implement full single-qubit operations, we still need an
operation which can make Bloch vector rotate around another axis
other than the z-axis. This operation can be achieved by changing
the magnetic flux\cite{Weichselbaum2004} as discussed in the
previous subsection. In the middle column of Fig.5, we have shown
such operation by changing the flux from $0.5\phi_0$ to
$0.445\phi_0$ from $t=0$. The electric field is set to zero. When
the rotation angle of Bloch vector is again equal to $\pi$, we can
get the NOT-gate which changes the state $|0\rangle$ to $|1\rangle$
and vice versa. The time for the NOT operation is $37.23ps$ which is
a little longer than the Z-gate but still much smaller than the
decoherence time.

With these two operations, we can implement arbitrary single-qubit
operations. For example, the Hadamard gate which may be the most
important single-qubit gate in quantum computation can be achieved
by accurately control the external fields and the operation time as
shown in the right column of Fig.5. With the pulses ${\bf
F}=0.00563{\text F}_0$ and $\phi=0.445\phi_0$, the time for a
Hadamard operation is $26.65ps$.

Besides the operation time, there is still another important index
of quantum gate. We know that there is a small fraction of
probability (P) which is lost from the first two states to the
higher-lying ones during every gate operation. This phenomenon will
result in errors in quantum computation and may also be considered
as a decoherence source. Then this probability must be regarded as
an important judgement of the validity of the qubit scheme.

In Fig.5 we have shown two kinds of probability loss corresponding
to different forms of the external field pulses. It can be seen that
the value of P can be less than $0.1\%$ if the pulse has a smooth
rising or trailing edge. Although such an error rate is still
greater than the threshold ($10^{-4}$ for an estimated value widely
accepted at present) for fault-tolerance quantum computation, it is
indeed small enough for coherence quantum operations in single-qubit
experiments. If the changes of external field parameters are
instantaneous, P will be much larger and have severe vibration for
both electric fields and magnetic flux. This also coincides with the
idea that the evolution of the qubit states should be adiabatic
during the quantum operations. A smooth change of the fields ensures
such premise.

\subsection{Structure and field effects}
\begin{figure}[ht]
\includegraphics*[angle=0,width=0.48\textwidth]{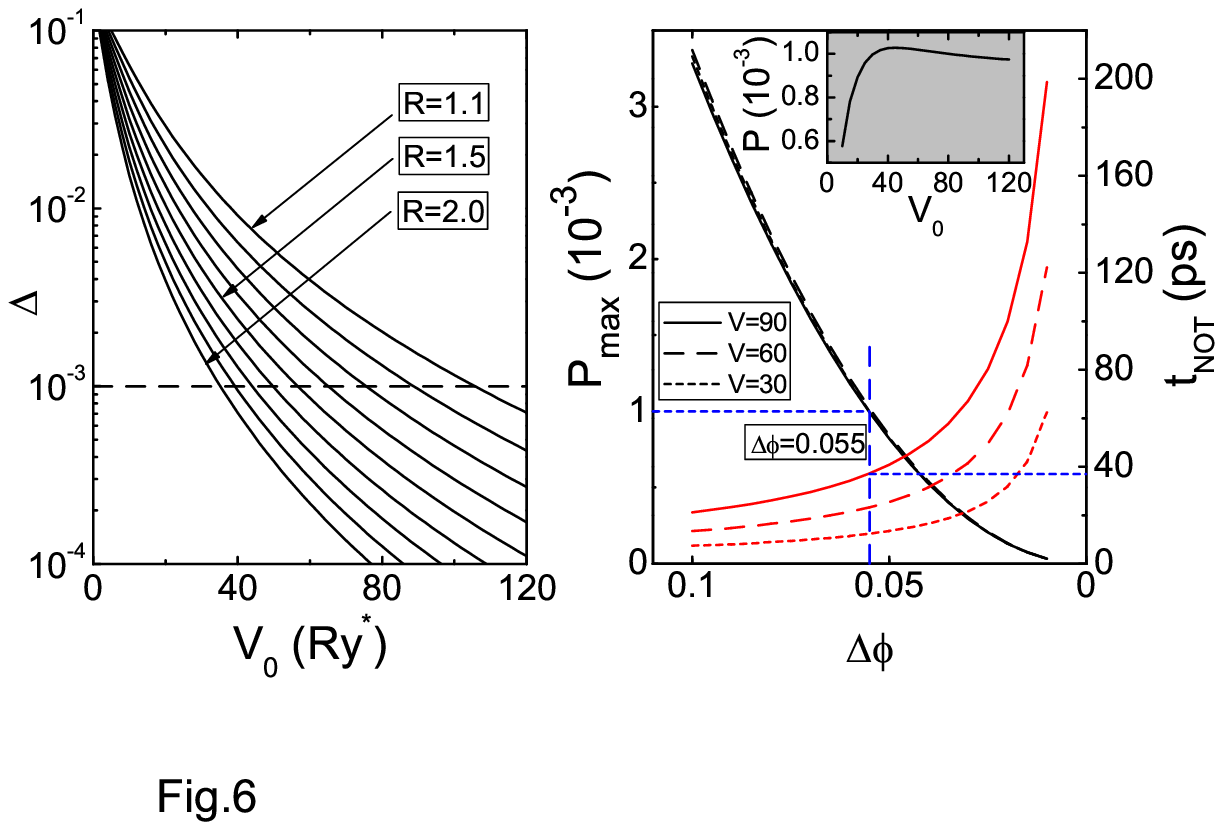}\\
\caption{\label{FIG:SizeEff}(Color online) (a) Error probability
($\Delta$) as functions of $V_0$ with different $R$. The curves
correspond to $R=1.1a_B^*$ to $R=2.0a_B^*$ from the top down.
(b)Maximal probability loss ($P_{max}$) and operation time
($t_{NOT}$) as functions of the change of magnetic flux
($\Delta\phi$) with three different $V_0$ (plotted with different
line styles) during the NOT-gate operation. The $V_0$-dependence of
P is shown in the inset. In all the cases, $W=0.4a_B^*$.}
\end{figure}
In this section, we will discuss the structure and field effects on
the validity of our qubit and corresponding operations.

First of all, we have chosen the two localized states of the
electron in a double-barrier nanoring as the logical states
$|0\rangle$ and $|1\rangle$. However, because of the finite height
of the two barriers, this two states are not completely localized in
one segment of the ring. The state mainly localized in one segment
indeed has a probability ($\Delta$) expanding to another segment.
This probability leads to errors in readout process and must be
limited in a tiny value. It can be convinced from Fig.6(a) that the
probability can be depressed by choosing appropriate $V_0$ for
different $R$. And the larger $R$ is, the lower $V_0$ is needed.

Second, we have seen that changing the magnetic flux can implement
the NOT-gate operation. But its operation time is longer than the
Z-gate which is implemented by applying a small electric field. Of
course, this operation time is related to the amplitude of the
change of the flux. It can be seen in Fig.6(b) that increasing the
change of the flux can apparently speed the operation. But on the
other hand, it also increases the probability loss during the
operation. So in our calculation we have chosen the field parameters
carefully to ensure both a short operation time and a small P. It is
worthwhile to indicate that the value of $V_0$ can also affect the
operation time and P. Lower barriers can speed the operation and
decrease P. But in fact it must be selected high enough to avoid the
increase of $\Delta$.

\begin{figure}[ht]
\includegraphics*[angle=0, width=0.35\textwidth]{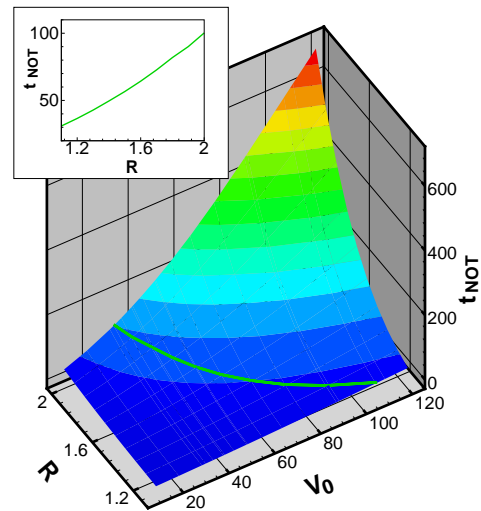}\\
\caption{\label{FIG:SizeEff}(Color online) Operation time for the
NOT-gate as a function of $R$ and $V_0$. An example of appropriate
choice of the structural parameters is illustrated by the green line
and its projection to the $R-t_{NOT}$ plane is plotted in the
inset.}
\end{figure}

Finally, we present $R$ and $V_0$-dependence of the operation time
for the NOT-gate in Fig.7. It can be found intuitively that large
radius $R$ and high barriers $V_0$ will dramatically increase the
operation time. An appropriate choice of the structural parameters
is illustrated by the green line in Fig.7 where small rings
correspond to high barriers and large rings correspond to low
barriers. Such selections can take into account both the speed of
the operation and tolerable errors.

\section{Summary}
A new scheme of qubit based on one electron's charge degree of
freedom in a double-barrier nanoring is presented. Because the first
two states of the system are degenerate when $\phi=0.5\phi_0$, the
logical states can be frozen in any linear superposition in the
qubit space to avoid unnecessary evolutions. The electric fields can
implement the z-axis gate operations of the qubit. By virtue of the
ringlike geometry of the system, the x-axis operations can be
achieved by AB effect of the magnetic flux. As a result, full qubit
operations can be implemented even if the barrier height is kept
constant. The structure and field effects are important for the
validity of the qubit. The external field pulse should also have an
appropriate rising or trailing edge to decrease the transition of
the electron to higher-lying states. The radius and barrier height
should be selected appropriately to speed the operations and depress
the errors. These results will be helpful in understanding the
evolutions of wavefunctions during the quantum operations and useful
for future implementation of qubits in solid system.

\begin{acknowledgments}
Financial supports from NSF-China (Grant No. 10374057 and 10574077) and ``973''
Programme of China (No. 2005CB623606) are gratefully acknowledged.
\end{acknowledgments}

\appendix*



\begin{thebibliography}{}
\bibitem{Kane1998} B. E. Kane, Nature {\bf 393}, 133 (1998).

\bibitem{Golovach2002} V. N. Golovach and D. Loss, Semicond. Sci. Technol. {\bf
17}, 355 (2002).

\bibitem{Elzerman2004} J. M. Elzerman, R. Hanson, L. H. W. van Beveren, B. Witkamp, L. M. K. Vandersypen and L. P. Kouwenhoven, Nature {\bf 430},
431 (2004).

\bibitem{Greentree2005} A. D. Greentree, A. R. Hamilton, L. C. L. Hollenberg and R. G. Clark, Phys. Rev. B {\bf
71}, 113310 (2005).

\bibitem{Hollenberg2004} L. C. L. Hollenberg, A. S. Dzurak, C. Wellard, A. R. Hamilton, D. J. Reilly, G. J. Milburn and R. G. Clark, Phys. Rev. B {\bf
69}, 113301 (2004).

\bibitem{Openov2004} L. A. Openov, Phys. Rev. B {\bf 70}, 233313 (2004).

\bibitem{Barrett2006} S. D. Barrett and T. M. Stace, Phys. Rev. Lett. {\bf
96}, 017405 (2006).

\bibitem{Hayashi2003} T. Hayashi, T. Fujisawa, H. D. Cheong, Y. H. Jeong and Y. Hirayama, Phys. Rev. Lett. {\bf
91}, 226804 (2003).

\bibitem{Gorman2005} J. Gorman, D. G. Hasko and D. A. Williams, Phys. Rev. Lett. {\bf
95}, 090502 (2005).

\bibitem{Fedichkin2000} L. Fedichkin, M. Yanchenko and K. A. Valiev, Nanotechnology {\bf
11}, 387 (2000).

\bibitem{Weichselbaum2004} A. Weichselbaum and S. E. Ulloa, Phys. Rev. A {\bf
70}, 032328 (2004).

\bibitem{Ulloa2004} A. Weichselbaum and S. E. Ulloa, Phys. Rev. B {\bf
70}, 195332 (2004).

\bibitem{Lorke2000} A. Lorke, R. J. Luyken, A. O. Govorov, J. P. Kotthaus, J. M. Garcia and P. M. Petroff, Phys. Rev. Lett. {\bf
84}, 2223 (2000).

\bibitem{Fuhrer2001} A. Fuhrer, S. L\"{u}scher, T. Ihn, T. Heinzel, K. Ensslin, W. Wegscheider and M. Bichler, Nature. {\bf 413}, 822 (2001).

\bibitem{Keyser2002} U. F. Keyser, S. Borck, R. J. Haug, M. Bichler, G. Abstreiter and W.
Wegscheider, Semicond. Sci. Technol. {\bf 17}, L22 (2002).

\bibitem{Granados2003} D. Granados and J. M. Garc\'{i}a, Appl. Phys. Lett. {\bf 82}, 2401 (2003).

\bibitem{Hobbs2004} K. L. Hobbs, P. R. Larson, G. D. Lian, J. C. Keay and M. B. Johnson, Nano Lett. {\bf 4}(1), 167 (2004).

\bibitem{JHHe2005} J. H. He, W. W. Wu, Y. L. Chueh, C. L. Hsin, L. J. Chen and L. J. Chou, Appl. Phys. Lett. {\bf 87}, 223102 (2005).

\bibitem{Jiang2003} N. Jiang, G. G. Hembree, J. C. H. Spence, J. Qiu, F. J. Garcia de Abajo and J. Silcox, Appl. Phys. Lett. {\bf 83}, 551 (2003).

\bibitem{JHHe2004} J. H. He, Y. L. Chueh, W. W. Wu, S. W. Lee, L. J. Chen and L. J. Chou, Thin Solid Films {\bf 469}, 478 (2004).

\bibitem{Kong2004} X. Y Kong, Y. Ding, R. Yang, and Z. L. Wang, Science {\bf
303}, 1348 (2004).

\bibitem{Zhu2003} J. -L. Zhu, X. Q. Yu, Z. H. Dai and X. Hu, Phys. Rev. B {\bf
67}, 075404 (2003).

\bibitem{Cedraschi2001} P. Cedraschi and M. B\"{u}tiker, Ann. Phys. {\bf
289}, 1 (2001).

\bibitem{Buttiker2005} M. B\"{u}ttiker and A. N. Jordan, Physica E {\bf
29}, 272 (2005).

\bibitem{Buttiker1996} M. B\"{u}ttiker and C. A. Stafford, Phys. Rev. Lett. {\bf
76}, 495 (1996).

\bibitem{Aassime2001} A. Aassime, G. Johansson, G. Wendin, R. J. Schoelkopf and P. Delsing, Phys. Rev. Lett. {\bf
86}, 3376 (2001).

\bibitem{Buks1998} E. Buks, R. Schuster, M. Heiblum, D. Mahalu and V. Umansky, Nature {\bf
391}, 871 (1998).

\bibitem{Pilgram2002} S. Pilgram and M. B\"{u}ttiker, Phys. Rev.
Lett. {\bf 89}, 200401 (2002)

\bibitem{Gurvitz2003} S. A. Gurvitz, L. Fedichkin, D. Mozyrsky, and G. P. Berman, Phys. Rev. Lett. {\bf
91}, 066801 (2003).

\end{thebibliography}
\end{document}